\documentclass[nohyperref]{article}
\pdfoutput=1

\usepackage{aas_macros}

\usepackage{microtype}
\usepackage{graphicx}
\usepackage{subcaption}
\usepackage{booktabs} 

\usepackage{hyperref}

 \usepackage[accepted]{icml2022}

\usepackage{amsmath}
\usepackage{amssymb}
\usepackage{mathtools}
\usepackage{amsthm}
\usepackage{enumitem}
\usepackage{bm, bbm}
\usepackage{lipsum}
\usepackage{physics}
\DeclareRobustCommand{\bbone}{\text{\usefont{U}{bbold}{m}{n}1}}

\usepackage[capitalize,noabbrev]{cleveref}

\theoremstyle{plain}

\theoremstyle{definition}

\theoremstyle{remark}

\usepackage[textsize=tiny]{todonotes}

\icmltitlerunning{Pixelated Reconstruction of Gravitational Lenses}

\begin{document}

\twocolumn[
\icmltitle{Pixelated Reconstruction of Gravitational Lenses using Recurrent Inference Machines}



\icmlsetsymbol{equal}{*}

\begin{icmlauthorlist}
\icmlauthor{Alexandre Adam}{udm,mila}
\icmlauthor{Laurence Perreault-Levasseur}{udm,mila,cca}
\icmlauthor{Yashar Hezaveh}{udm,cca}
\end{icmlauthorlist}

\icmlaffiliation{udm}{Department of Physics, University of Montreal, Canada}
\icmlaffiliation{mila}{Mila --- Quebec Artificial Intelligence Institute, Montreal, Canada}
\icmlaffiliation{cca}{Center for Computational Astrophysics, Flatiron Institute, New-York, USA}

\icmlcorrespondingauthor{Alexandre Adam}{alexandre.adam@umontreal.ca}
\icmlcorrespondingauthor{Laurence Perreault-Levasseur}{laurence.perreault.levasseur@umontreal.ca}

\icmlkeywords{Machine Learning, ICML}

\vskip 0.3in
]



\printAffiliationsAndNotice{}  

\begin{abstract}
Modeling strong gravitational lenses in order to quantify the distortions in the images of background sources and to reconstruct the mass density in the foreground lenses has traditionally been a difficult computational challenge. As the quality of gravitational lens images increases, the task of fully exploiting the information they contain becomes computationally and algorithmically more difficult. In this work, we use a neural network based on the Recurrent Inference Machine (RIM) to simultaneously reconstruct an undistorted image of the background source and the lens mass density distribution as pixelated maps. The method we present iteratively reconstructs the model parameters (the source and density map pixels) by learning the process of optimization of their likelihood given the data using the physical model (a ray-tracing simulation), regularized by a prior implicitly learned by the neural network through its training data. When compared to more traditional parametric models, the proposed method is significantly more expressive and can reconstruct complex mass distributions, which we demonstrate by using realistic lensing galaxies taken from the cosmological hydrodynamic simulation IllustrisTNG.

\end{abstract}
\section{Introduction}
Strong gravitational lensing is a natural phenomenon through which multiple distorted images of luminous background objects, i.e. early-type star-forming galaxies, are formed by massive foreground objects along the line of sight \citep[e.g.,][]{Viera2013,Marrone2018,Rizzo2020,Sun2021}. 
These distortions are tracers of the distribution of mass in foreground objects, independent of the electromagnetic behaviour of these overdensities. 
As such, this phenomenon offers a powerful probe of the distribution of dark matter and its properties outside of the Milky Way \citep[e.g.,][]{Dala2002,Treu2004,Hezaveh2016,Gilman2020,Gilman2021}.

Lens modeling is the process of inferring the parameters describing both the mass distribution in the foreground lens and the light emitted by the background source. This has traditionally been a time- and resource-consuming procedure. A common practice to model the mass of lensing galaxies is 
to assume that the density profiles 
follow simple parametric forms, e.g., a power law $\rho \propto r^{-\gamma'}$. 
These profiles generally provide a good fit to low-resolution data and are easy to work with due to their small number of parameters \citep[e.g.,][]{Koopman2006,Barnabe2009,Auger2010}. 
However, as high-resolution and high signal-to-noise ratio (SNR) images become available, lens analysis with simple models requires introducing additional parameters representing the complexities in the lensing galaxies and their immediate environments \citep[e.g.,][]{Sluse2017,Wong2017,Birrer2019,Rusu2019, Rusu2017,Li2021}. 
This approach becomes intractable as the quality of images increases. 
For example,
no simple parametric model of the Hubble Space Telescope (HST) Wide Field Camera 3 (WFC3) images of 
the Cosmic Horseshoe (J1148+1930) --- initially discovered by \citet{Belokurov2007} --- 
has been able to model the fine features of the extended arc 
\citep[e.g., ][]{Bellagamba2016,Cheng2019,Schuldt2019}.

In this work, we develop a method for pixelated 
strong gravitational lensing mass and source reconstruction, allowing it to reconstruct complex distributions. 
Our method is based on the Recurrent Inference Machine \citep[RIM,][]{Putzky2017}, 
which proposes to learn an iterative inference algorithm, moving away 
from hand-chosen inference algorithms and hand-crafted priors. 
In this framework, the prior is implicit in the dataset used to train 
the neural network. 
We also present a new architecture based on the original RIM to allow the inference of pixelated maps for this highly non-linear and under-constrained problem.

\section{Methods}\label{sec:methods}
\subsection{Data}\label{sec:data}
The background source brightness distributions are taken from the Hubble Space Telescope (HST) 
COSMOS field \citep{Koekemoer2007,Scoville2007}, acquired in the F814W filter. 
A dataset of mag limited ($\mathrm{F814W} < 23.5$) deblended 
galaxy postage stamps \citep{Leauthaud2007} was compiled as 
part of the GREAT3 challenge \citep{Mandelbaum2014}. The data is 
publicly available \citep{Mandelbaum2012}, and the preprocessing is done through 
the open-source software 
\texttt{GALSIM} \citep{Rowe2015}. The final set has 
$13\,321$ galaxy images cropped to $128^{2}$ pixels and with 
a flux greater than $50\,\,\mathrm{photons}\,\,\mathrm{cm}^{-2}\,\mathrm{s}^{-1}$. 
We split this set into a training set (90\%) and a test set (10\%) 
before data augmentation and denoising with an autoencoder 
\citep{Vincent2008}.

The projected surface density maps (convergence) of lensing galaxies 
were made using the redshift $z=0$ snapshot  
of the IllustrisTNG-100 simulation \citep{Nelson2018} 
in order to produce physically realistic realizations of dark matter and baryonic matter halos.
We selected 1604 profiles, split into a training set ($90\%$) and a test set ($10\%$), 
with the criteria that they have a total
dark matter mass of at least $9\times10^{11} M_{\odot}$. We then collected all 
dark matter, gas, stars, and black hole particles from the profiles. 
We then compute smoothed projected surface density maps with an adaptive Gaussian kernel
following the prescriptions from 
\citet{Auger2007} and \citet{Rau2013}. 
The final training set is composed of 3 different projections 
($xy$, $xz$ and $yz$) of each profiles, 
rendered on a pixelated grid with a resolution of $0.55\,\mathrm{kpc}/h$ and 
$128^{2}$ pixels.
Several data augmentation rounds, including rescaling 
each convergence maps randomly to produce a set with an Einstein radius 
uniformly distributed ${\theta_E \sim \mathcal{U}(0.5'',2.5'')}$, 
were used to increase the number of profiles to $50\,000$ 
for training a VAE and the RIM.

\subsection{Data Augmentation with VAEs}
When working with limited data, data augmentation is crucial to ensure that 
the trained model is robust against perturbations ---
like rotations of images --- 
which are not implicitly included as symmetries in the architecture of the model. 
We trained a variational auto-encoder \citep[VAE, ][]{Kingma2013} for data augmentation 
of the source maps and another VAE for the convergence maps.

Direct optimization 
of the ELBO loss for VAEs can prove difficult 
because the reconstruction term 
could be relatively weak compared to the Kullback Leibler (KL) divergence term 
\citep{Kingma2019}. To alleviate this issue, 
we follow the work of \citet{Bowman2015} and \citet{Sonderby2016} in setting a warm-up 
schedule for the KL term in the ELBO loss, 
starting from $\beta=0.1$ up to $\beta_{\mathrm{max}}$.
Following the work of \citet{Lanusse2021}, we also introduce 
an $\ell_{2}$ penalty between the input and output of the bottleneck 
fully-connected layers to encourage an identity map between them. This regularisation 
term is slowly removed during training.

\begin{figure}[tb!]
       \centering 
       \includegraphics[width=\linewidth]{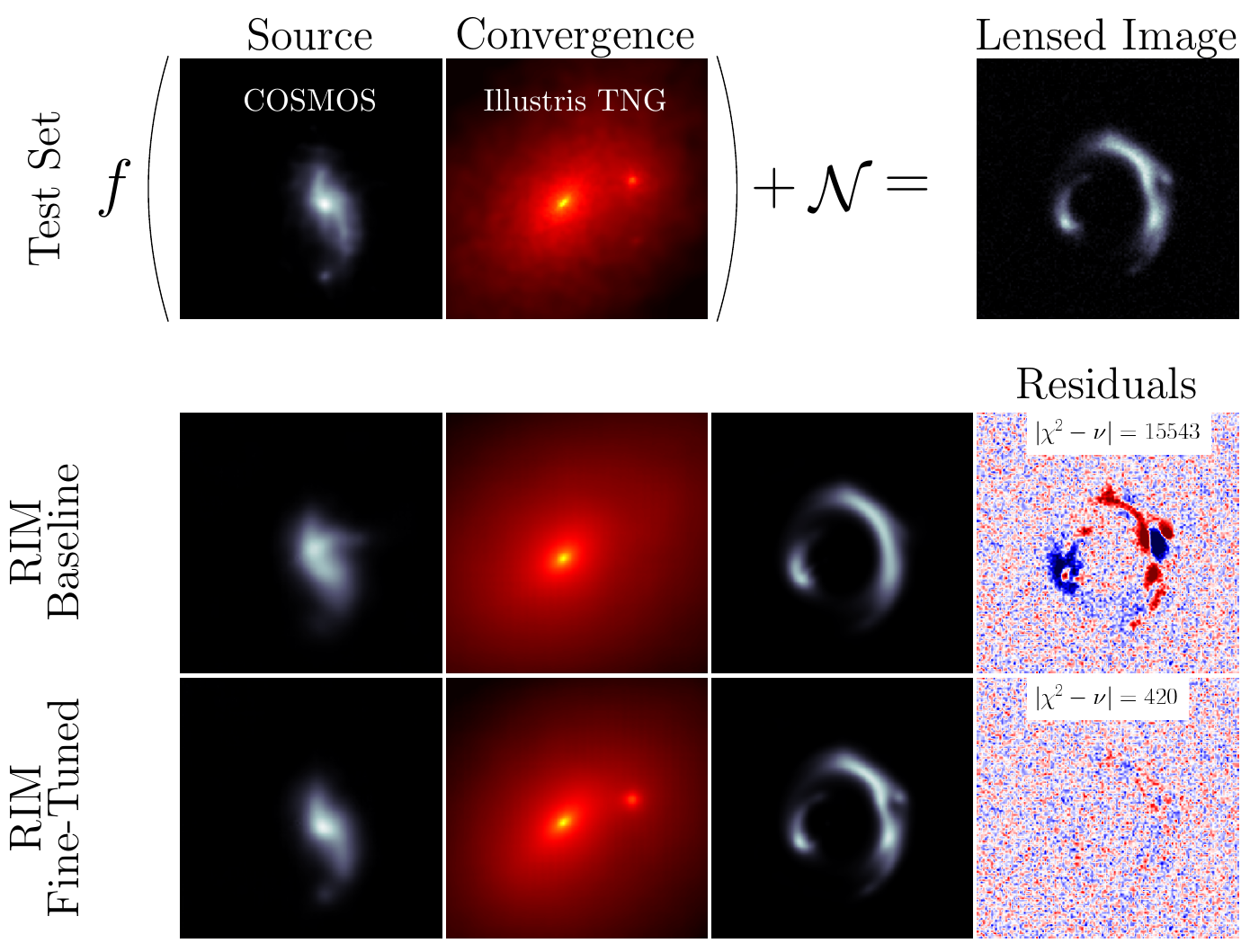}
       \caption{Example of a simulated lensed image in the test set that 
exhibits a large deflection in its eastern arc which indicates the presence of a massive object
 --- in this case a dark matter subhalo. The fine-tuning procedure is able to recover 
this subhalo because of its strong signal in the lensed image and reduces the residuals 
to noise level.}
       \label{fig:main figure}
\end{figure}

\subsection{Raytracing}
Simulations of lensed images are produced using a ray-tracing code, which maps the brightness distribution
of background sources to the observed coordinates. 
The foreground pixel coordinates $\boldsymbol{\theta}_i$ and the source 
pixel coordinates $\boldsymbol{\beta}_i$ are related by
the lens equation
\begin{equation}\label{eq:LensEquation}
        \bm{\beta}_i = \bm{\theta}_i - \bm{\alpha}(\bm{\theta}_i),
\end{equation}
where $\boldsymbol{\alpha}$ is the deflection angle.
The deflection angle is calculated from the projected surface 
density field $\kappa$ (also commonly refered to as the convergence) by the integral
\begin{equation}\label{eq:alpha}
        \bm{\alpha}(\boldsymbol{\theta}_i) = 
        \frac{1}{\pi} \int_{\mathbb{R}^2}
        \kappa(\boldsymbol{\theta}') 
        \frac{\boldsymbol{\theta}_i
        - \boldsymbol{\theta}'}{\lVert \boldsymbol{\theta}_i - 
        \boldsymbol{\theta}' \rVert^2}
        d^2\boldsymbol{\theta}'
\end{equation}
The intensity of a pixel in a simulated lensed image 
is obtained by bilinear interpolation of the 
source brightness distribution at the coordinate $\boldsymbol{\beta}_i$.
The integral 
in equation \eqref{eq:alpha} 
is computed in near-linear time using Fast Fourier Transforms (FFT). 

A blurring operator --- i.e., a convolution by a point spread function --- is then 
applied to the lensed image to replicate the response of an imaging system. 
This operator is implemented as a GPU-accelerated matrix operation 
since the blurring kernels used in this paper have a significant proportion
of their energy distribution encircled inside a small pixel radius.

\begin{figure}
        \centering
        \includegraphics[width=\linewidth]{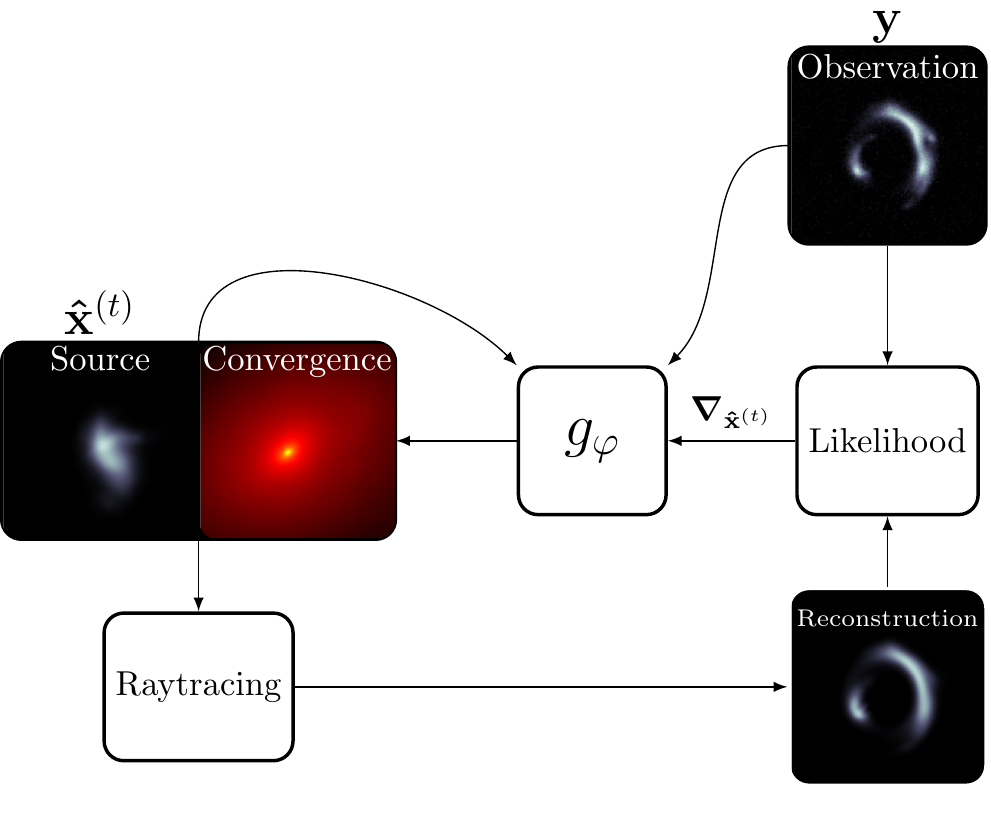}
        \caption{Rolled computational graph of the RIM.}
        \label{fig:rolled_graph}
\end{figure}

\begin{figure*}[ht!]
        \centering
        \includegraphics[width=\textwidth]{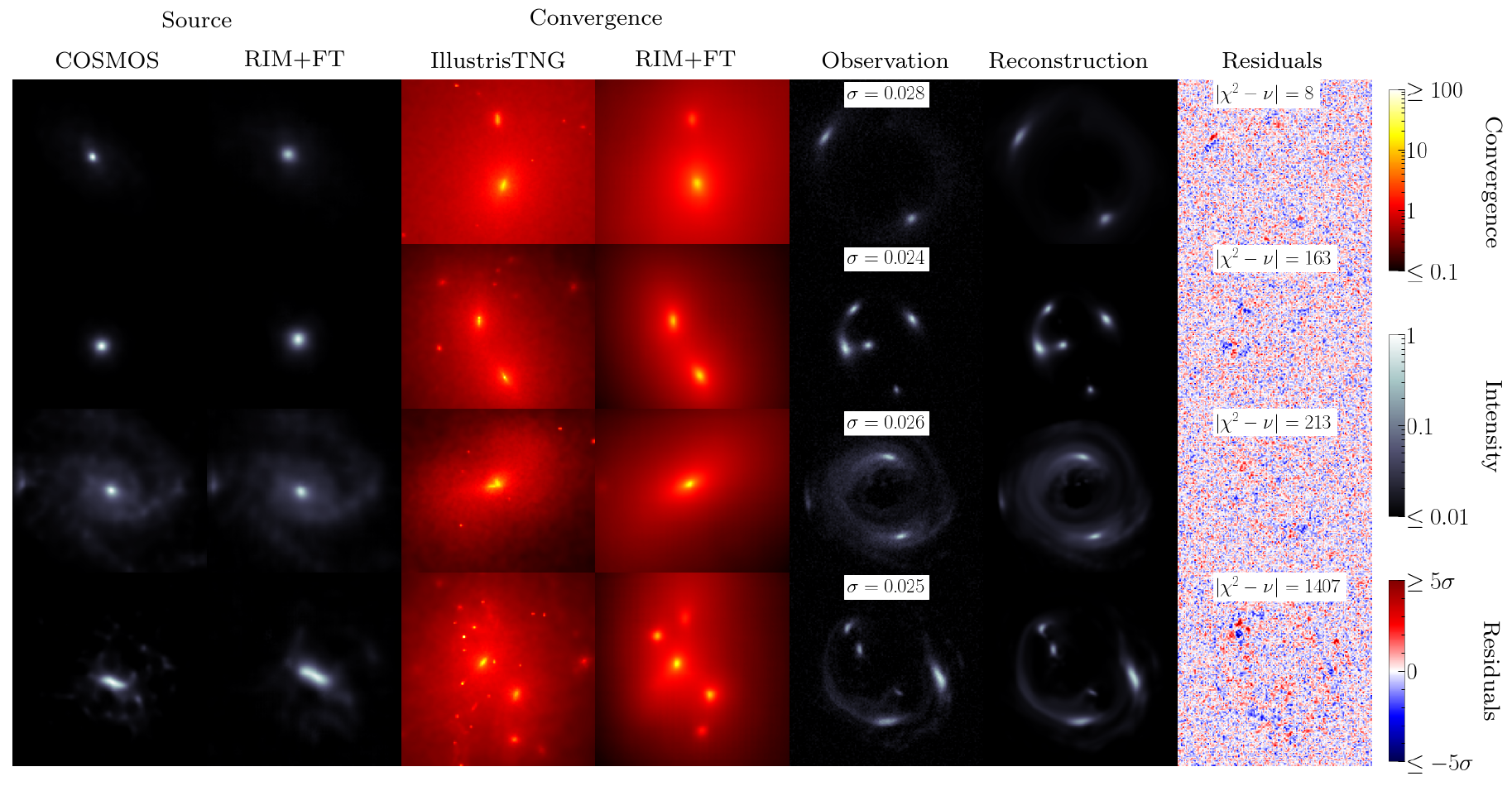}
        \caption{
                Samples of the fine-tuned RIM (RIM+FT) reconstructions 
                from the test set. 
                We report the residuals of the reconstruction, as well as the difference between the $\chi^2$ and the number of degrees of freedom $\nu = 16384$.
} 
        \label{fig:main result}

        \newpage
\end{figure*}

\subsection{Recurrent Inference Machine}
The RIM \citep{Putzky2017} is a form of learned 
gradient-based inference intended to solve inverse problems of the form 
\begin{equation}
\mathbf{y} = f(\mathbf{x}) + \mathcal{N}\,, 
\end{equation}
where $\mathbf{y}$ is a vector of noisy lensed images, $f$ is a function encoding the physical model, $\mathbf{x}$ is a vector of parameters of interest, and $\mathcal{N}$ is a vector of additive noise. This framework has been applied in the context of linear inverse problems, where the function $f$ can be represented in a matrix form, in particular in the cases of under-constrained problems
for which the prior on the parameters $\mathbf{x}$, $p(\mathbf{x})$, is 
either intractable or hard to compute \citep{Morningstar2018,Morningstar2019,Lonning2019}. 
The use of the RIM to solve non-linear inverse problems was first investigated in \citep{Modi2021}.
In our case, the inverse problem mapping function $F$ is non-linear w.r.t. the convergence parameters which is a consequence of the non-linearity of equation \eqref{eq:LensEquation}. 

The governing equation for the RIM is a recurrent 
relation that takes the general form
\begin{equation}\label{eq:RIM} 
        \mathbf{\hat{x}}^{(t+1)} = \mathbf{\hat{x}}^{(t)} 
        + g_\varphi \big(\mathbf{\hat{x}}^{(t)},\, \mathbf{y},\, 
\grad_{\mathbf{\hat{x}^{(t)}}} \log p(\mathbf{y} \mid \mathbf{x}^{(t)})\big)\, ,
\end{equation}
where $p(\mathbf{y} \mid \mathbf{x})$ is an isotropic gaussian likelihood function, characterized 
by the noise standard deviation $\sigma$.
By minimizing the weighted mean squared loss
\begin{equation}\label{eq:Loss}
		\mathcal{L}_\varphi(\mathbf{x}, \mathbf{y}) = \frac{1}{T}\sum_{t=1}^{T}\sum_{i=1}^{M} \mathbf{w}_i (\mathbf{\hat{x}}^{(t)}_i - \mathbf{x}_i)^2\, ,
\end{equation} 
the RIM learns to optimize the 
parameters $\mathbf{x}$ given a likelihood function. 
Unlike previous work 
\citep{Andrychowicz2016,Putzky2017,Morningstar2018,Morningstar2019,Lonning2019}, 
the data vector $\mathbf{y}$ --- or observation --- 
is fed to the neural network in order to learn 
the initialization of the parameters, $\mathbf{\hat{x}}^{(0)} = g_\varphi(0, \mathbf{y}, 0)$, 
as well as their optimization. 
We found in practice that this significantly improves the performance 
of the model for our problem and it avoids situations where 
the model would get stuck in local minima at 
test time due to poor initialization. 

The RIM used in this work is designed based on a U-net architecture \citep{Ronneberger2015}. 
The most important aspect of our implementation is the use of
Gated Recurrent Units \citep[][]{Cho2014} placed in each skip connection 
which guides the reconstruction independently at different levels 
of resolution. 
The gradient of the likelihood is computed using automatic differentiation. Following 
\citet{Modi2021}, we preprocess the gradients using the Adam algorithm \citep{Kingma2013}.

\subsection{Fine-tuning}
Once the RIM is trained, 
we can treat the RIM optimization procedure as a baseline 
estimator of the parameters 
$\mathbf{x}$ given a noisy observation $\mathbf{y}$. 
We now concern ourselves with a strategy to improve 
this estimator. 
This is important 
for observations with high SNR, for which the estimator
must be extremely accurate to model all the  fine features present in the arcs.
The fine-tuning objective is to minimize directly   
the likelihood over each time steps of the RIM:
\begin{equation}\label{eq:MAP} 
        \hat{\varphi}_{\mathrm{MAP}} = \underset{\varphi}{\mathrm{argmax}}\,\, 
        \frac{1}{T}\sum_{t=1}^{T} \log p(\mathbf{y} \mid \mathbf{\hat{x}}^{(t)}) + \log p(\varphi)\, .
\end{equation} 
This objective function makes no use of labels, meaning that only an observation $\mathbf{y}$ 
is required to fine-tune the RIM.
This allows us to use this objective at test time, at which point the 
RIM is trained to reconstruct this specific observation. 

We use elastic weight consolidation \citep[EWC, ][]{Kirkpatrick2016} as 
prior over the model parameters. 
To compute the Fisher matrix in EWC, it is necessary to sample from a distribution of lensing systems that are conditioned on the observation. 
This is accomplished by sampling the latent space of both the source VAE and the convergence VAE near the latent code 
of the baseline prediction of the RIM. More details regarding this procedure are given in appendix 
\ref{ap:fine tuning}.

\section{Results and Discussion}\label{sec:results}

Figure \ref{fig:main result} presents a few examples of the reconstructions obtained using the approach presented above. An emphasis is put on complex convergence profiles with multiple 
main deflectors or substructures. Modeling such convergence maps with traditional maximum-likelihood methods using analytical profiles would require significant user input and considerable computational resources due to large parameter degeneracies.
The mock observations as well as the reconstructed 
lensed images are also shown, 
alongside the residuals of the reconstructions and the $\chi^2$ 
statistic. The first 3 reconstructions have statistically significant 
residuals, with no pixels exceeding the $5 \sigma$ threshold. The last 
reconstruction, which is arguably the most complex to perform, 
has a few pixels reaching $5\sigma$. 

\begin{figure}[t!]
        \centering
        \includegraphics[width=\linewidth]{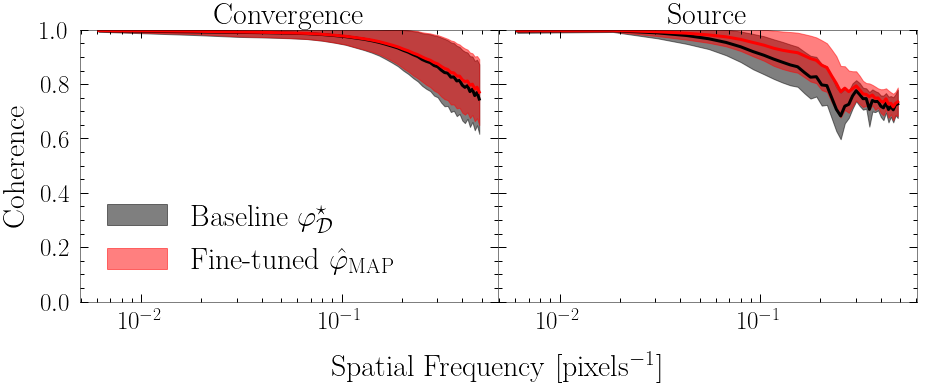}
        \caption{Statistics of the coherence spectrum from a test set. 
		The solid line is the average 
        coherence. The transparent region is the $68\%$ confidence interval. 
}
        \label{fig:coherence}
\end{figure}

In addition to a visual inspection of the reconstructed sources 
and convergences, we compute 
the coherence spectrum to quantitatively assess the quality the reconstructions
\begin{equation}\label{eq:coherence} 
        \gamma(k) = \frac{P_{12}(k)}{\sqrt{P_{11}(k) P_{22}(k)}} \, ,
\end{equation}
where $P_{ij}(k)$ is the cross power spectrum of images $i$ and $j$ at 
the wavenumber $k$. Figure \ref{fig:coherence} shows the mean value and the $68\%$ inclusion interval of those
spectra
for the convergence and source maps in a test set of 3000 examples. 
The fine-tuning 
procedure, shown in red, is able to improve significantly the coherence of the baseline background 
source, shown in black, at all scales. The coherence spectrum of the convergence remains unchanged by the fine-tuning procedure. 
Still, we note that many examples in the dataset showcase significant 
improvement which we illustrate in Figure \ref{fig:main figure}.

The results obtained here demonstrate the effectiveness of machine learning methods for inferring pixelated maps of the distribution of mass in lensing galaxies. 
Since this is a heavily under-constrained problem, stringent priors are needed to avoid overfitting the data, a task that has traditionally been difficult to accomplish \citep[e.g., ][]{Saha1997}. 
The model proposed here can implicitly learn these priors from a set of training data.

The flexible and expressive form of the reconstructions means that, in principle, any lensing system (e.g., a single simple galaxy, or a group of complex galaxies) could be analyzed by this model, without any need for pre-determining the model parameterization. This is of high value given the diversity of observed lensing systems, and their relevance for constraining astrophysical and cosmological parameters. 


\section*{Software and data}
The source code, as well as the various scripts and parameters used to 
produce the model and results is available as open-source software 
under the package \texttt{Censai}\footnote{\href{https://github.com/AlexandreAdam/Censai}{
\includegraphics[scale=0.25]{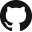}
https://github.com/AlexandreAdam/Censai}}. 
The model parameters, as well as the convergence maps and the background sources used to train 
these models, the test set examples and the reconstructions results are available as open-source datasets hosted by Zenodo\footnote{\href{https://doi.org/10.5281/zenodo.6555463}
{\includegraphics[scale=0.1]{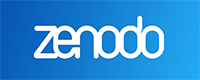}
https://doi.org/10.5281/zenodo.6555463}}. 
This research made use of \texttt{Tensorflow} \citep{tensorflow}, \texttt{Tensorflow-Probability} \citep{tensorflow-probability}, \texttt{Numpy} \citep{numpy}, \texttt{Scipy} \citep{scipy}, \texttt{Matplotlib} \citep{matplotlib}, \texttt{scikit-image} \citep{scikit-image}, \texttt{IPython} \citep{ipython}, \texttt{Pandas} \citep{pandas1,pandas2}, \texttt{Scikit-learn} \citep{scikit-learn}, \texttt{Astropy} \citep{astropy:2013,astropy:2018} and \texttt{GalSim} \citep{galsim}.

\section*{Acknowledgements}
This research was supported by the Schmidt Futures Foundation. The work was also enabled in part by computational resources provided by Calcul Quebec, Compute Canada and the Digital Research Alliance of Canada. Y.H. and L.P. acknowledge support from the National Sciences and Engineering Council of Canada grant RGPIN-2020-05102, the Fonds de recherche du Québec grant 2022-NC-301305, and the Canada Research Chairs Program. A.A. was supported by an IVADO scholarship.

\bibliography{bibliography}
\bibliographystyle{icml2022}

\newpage
\onecolumn
\appendix

\begin{figure}[t]
        \centering
        \includegraphics[width=\textwidth]{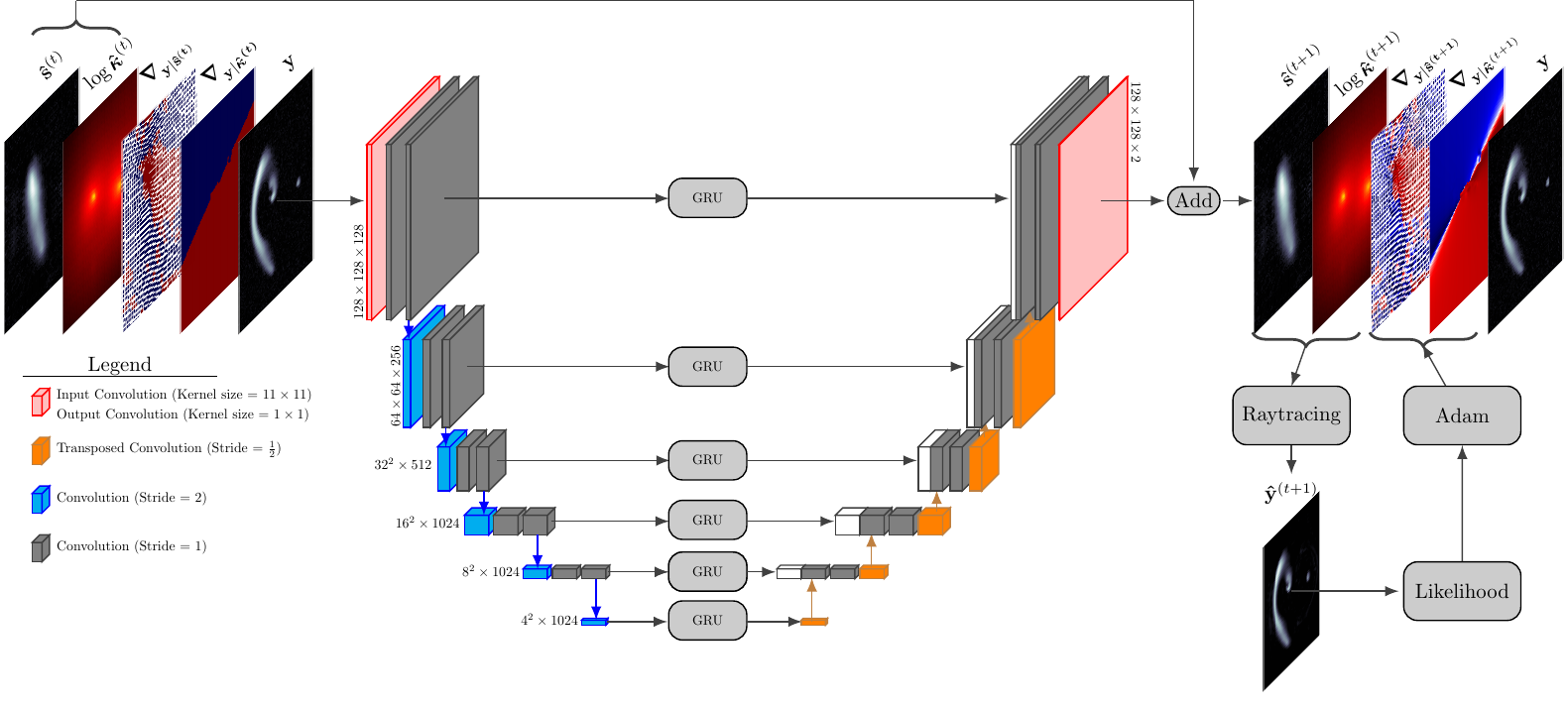}
        \caption{
A single time step of the unrolled computation graph of the RIM.
GRU units are placed in the skip connections to guide the 
reconstruction of the source and convergence. A diagram of the steps to compute 
the likelihood gradients is shown in the bottom right of the figure, including the 
Adam processing step. We used the shorthand notation $\grad_{\mathbf{y} \mid \mathbf{x}} \equiv 
\grad_{\mathbf{x}} \log p(\mathbf{y} \mid \mathbf{x})$, $\mathbf{s}$ to mean 
the source map and $\log \boldsymbol{ \kappa} $ to mean the $\log_{10}$ of 
the convergence map.
}
        \label{fig:unet}
\end{figure}

\section{Training dataset}

$400\,000$ observations are simulated from random pairs of COSMOS sources 
and IllustrisTNG convergence training splits in order to train the RIM. 
An additional $200\,000$ observations are created from pairs 
of COSMOS source and pixelated singular isothermal elliptical (SIE) convergence maps.
$1\,600\,000$ simulated observations are also generated from the VAE 
background sources and convergence maps as part of the training set. 
Validation checks are applied to each 
examples in order to avoid configurations like a
single image of the background 
source or an Einstein ring cropped by the field of view. 
\begin{table}[H]
		\centering
		\caption{SIE parameters.}
		\label{tab:sie}
		\begin{tabular}{ccc}
				Parameter &  Distribution \\
				\hline \hline
				 Radial shift ('') & $\mathcal{U}(0, 0.1)$ \\
				Azimutal shift & $\mathcal{U}(0, 2\pi)$ \\
				Orientation & $\mathcal{U}(0, \pi)$ \\
				$\theta_E$ ('') & $\mathcal{U}(0.5, 2.5)$ \\
				Ellipticity & $\mathcal{U}(0, 0.6)$ \\
				\hline
		\end{tabular}
\end{table}


\section{Data augmentation with VAE}

\begin{figure}[H]
		\centering
        \begin{subfigure}[t]{0.45\linewidth}
		        \includegraphics[width=0.9\linewidth]{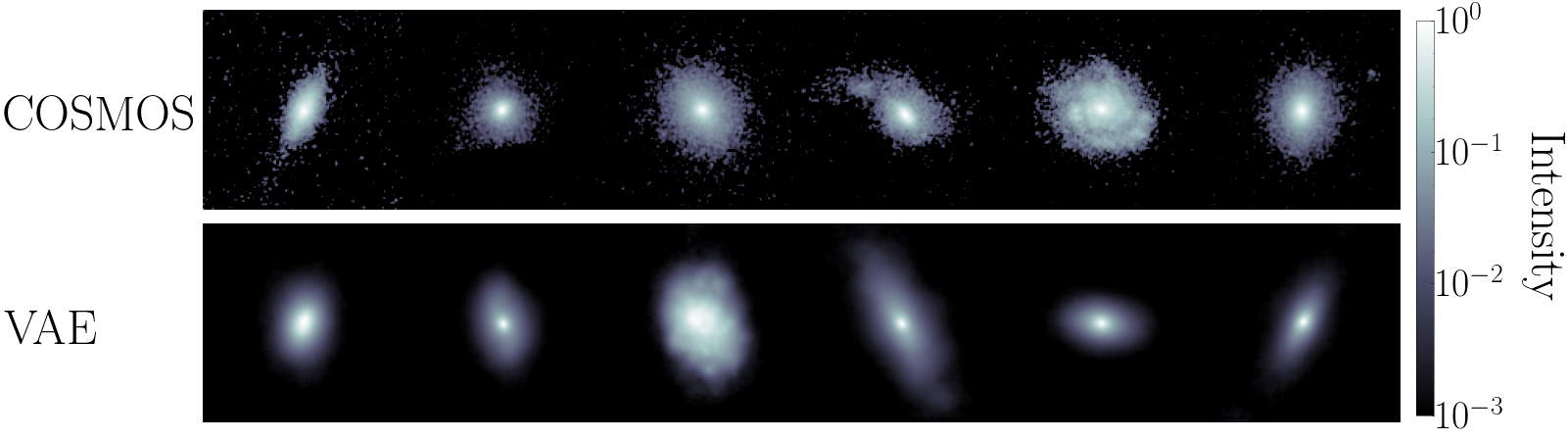}
                \caption{Examples of COSMOS galaxy images 
                        (top row) and VAE generated samples (bottom row).}
                \label{fig:source}
        \end{subfigure}
        ~
        \begin{subfigure}[t]{0.45\linewidth}
                \includegraphics[width=0.9\linewidth]{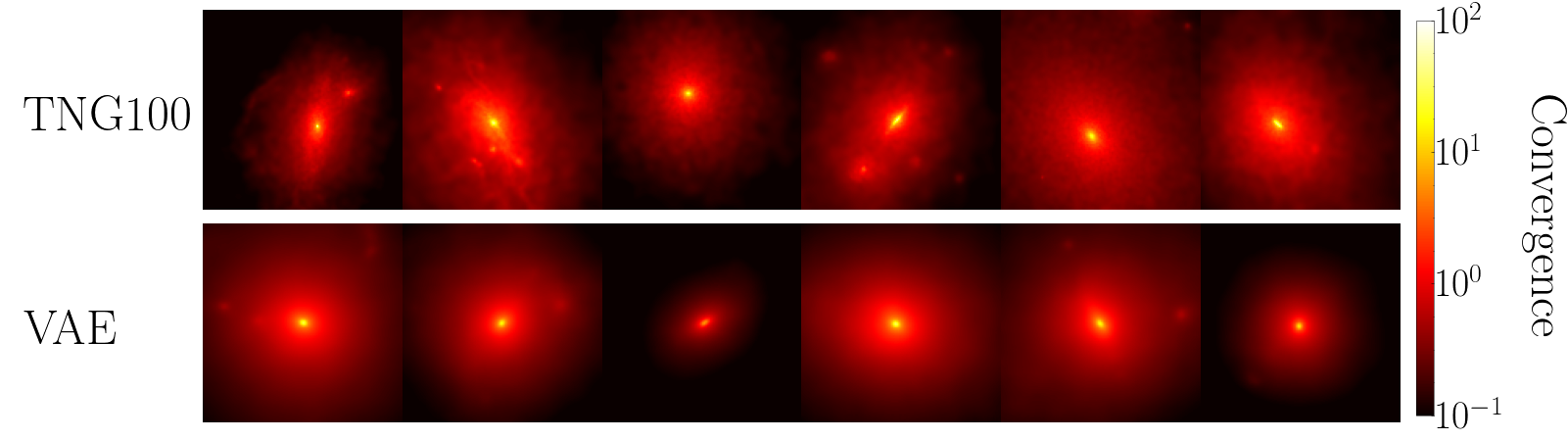}
                \caption{Examples of smoothed Illustris TNG100 convergence map (top row) 
                and VAE generated samples (bottom row).}
                \label{fig:kappa}
        \end{subfigure}
        \caption{Labels for the training set.}
\end{figure}

\section{VAE Architecture and optimisation}

For the following architectures, we employ the notion of \textit{level} 
to mean layers in the encoder and the decoder with the same spatial resolution. 
In each level, we place a block of convolutional layers 
before downsampling (encoder) or after upsampling (decoder). These operations 
are done with strided convolutions like in the U-net architecture of the RIM.

\begin{table}[H]
\begin{minipage}{.5\linewidth} 
        \centering
        \caption{Hyperparameters for the background source VAE.}
        \label{tab:Source VAE}
        \begin{tabular}{cc}
                Parameter & Value \\\hline\hline
                Input preprocessing & $\bbone$ \\
                                    & \\

                \textit{Architecture} & \\
                Levels (encoder and decoder) & 3 \\
                Convolutional layer per level & 2 \\
                Latent space dimension & 32\\
                Hidden Activations & Leaky ReLU \\
                Output Activation & Sigmoid \\
                Filters  & 16, 32, 64 \\
                Number of parameters & $3\,567\,361$\\

                           & \\
                \textit{Optimization} & \\
                Optimizer & Adam \\
                Initial learning rate & $10^{-4}$ \\
                Learning rate schedule & Exponential Decay \\
                Decay rate & 0.5 \\
                Decay steps & $30\,000$ \\
                Number of steps & $500\,000$ \\
                $\beta_{\mathrm{max}}$ & 0.1 \\
                Batch size & 20\\
                \hline
        \end{tabular}
\end{minipage}
\begin{minipage}{.5\linewidth}
        \caption{Hyperparameters for the convergence VAE.}
        \label{tab:Kappa VAE}
        \begin{tabular}{cc}
                Parameter & Value \\\hline\hline
                Input preprocessing & $\log_{10}$ \\
                              & \\

                \textit{Architecture} & \\
                Levels (encoder and decoder) & 4 \\
                Convolutional layer per level & 1 \\
                Latent space dimension & 16\\
                Hidden Activations & Leaky ReLU \\
                Output Activation & $\bbone$ \\
                Filters & 16, 32, 64, 128 \\
                Number of parameters & $1\,980\,033$\\

                           & \\
                \textit{Optimization} & \\
                Optimizer & Adam\\
                Initial learning rate & $10^{-4}$ \\
                Learning rate schedule & Exponential Decay \\
                Decay rate & 0.7 \\
                Decay steps & $20\,000$ \\
                Number of steps & $155\,000$ \\
                $\beta_{\mathrm{max}}$ & 0.2 \\
                Batch size & 32\\
                \hline
        \end{tabular}
\end{minipage}
\end{table}

\section{RIM architecture and optimisation}

\begin{figure}[H]
        \centering
        \includegraphics[width=\linewidth]{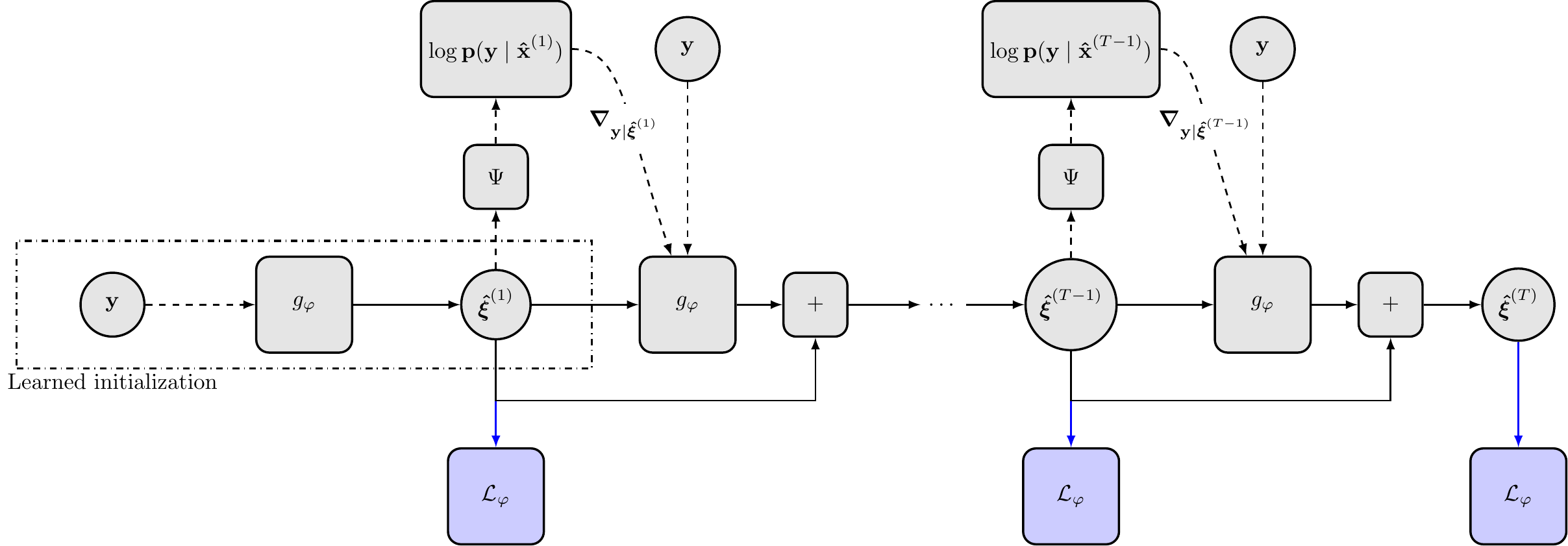}
        \caption{Unrolled computational graph of the RIM. Operations along solid arrows are being 
        recorded for BPTT, while operations along dashed arrows are not. The blue arrows are only 
        used for optimisation during training. During fine-tuning or testing, the loss is computed only 
        as an oracle metric to validate that our methods can recover the ground truth.}
        \label{fig:unrolled graph}
\end{figure}

The notion of link function $\Psi: \Xi \rightarrow \mathcal{X}$, 
introduced by \citet{Putzky2017}, is an invertible transformation 
between the network prediction space $\boldsymbol{\xi} \in \Xi$ 
and the forward modelling space $\mathbf{x} \in \mathcal{X}$.
This is a different notion from preprocessing, discussed in section \ref{sec:data}, 
because this transformation is applied inside the recurrent relation \ref{eq:RIM} 
as opposed to before training. In the case where the forward model has some restricted 
support or it is found that some transformation helps the training, then 
the link function chosen must be implemented as part of the network architecture as 
shown in the unrolled computational graph in Figure \ref{fig:unrolled graph}.
Also, the loss $\mathcal{L}_\varphi$ must be computed in the $\Xi$ space in order 
to avoid gradient vanishing problems when $\Psi$ is a non-linear mapping, which 
happens if the non-linear link function is applied in an 
operation recorded for backpropagation through time (BPTT).

For the convergence, we use an exponential link function with base $10$: 
$\boldsymbol{\hat{\kappa}} = \Psi(\boldsymbol{\xi}) = 10^{\boldsymbol{\xi}}$. 
This $\Psi$ encodes the non-negativity of the convergence. Furthermore, 
it is a power transformation that leaves the linked 
pixel values $\boldsymbol{\xi}_i$ normally distributed, thus improving the 
learning through the non-linearities in the neural network.
The pixel weights $\mathbf{w}_i$ in the loss function \eqref{eq:Loss}
are chosen to encode the fact that the pixel with critical mass density ($\boldsymbol{\kappa}_i > 1$) 
have a stronger effect on the lensing configuration than other pixels. 
We find in practice that the weights 
\begin{equation}\label{eq:convergence weights} 
        \mathbf{w}_i = \frac{\sqrt{\boldsymbol{\kappa}_i}}{ \sum_i \boldsymbol{\kappa}_i}, 
\end{equation} 
encode this knowledge in the loss function and improved both the empirical 
risk and the goodness of fit of the baseline model on early test runs.

For the source, we found that we do not need a link function 
--- its performance is generally better compared to other link function we tried like sigmoid and 
power transforms --- and we found that the pixel weights can be taken to 
be uniform, i.e. $\mathbf{w}_i = \frac{1}{M}$.

In the first optimisation stage, we trained 24 different architectures from a small 
set of valid hyperparameters previously identified for approximately 4 days (wall time using a single Nvidia A100 gpu). 
Following this first stage, 4 architectures were deemed efficient enough 
to be trained for an additional 6 days. 

\begin{table}[H]
        \centering
        \caption{Hyperparameters for the RIM.}
        \label{tab:baseline hparams}
        \begin{tabular}{cc}
                Parameter & Value \\\hline\hline
                Source link function & $\bbone$ \\
                $\kappa$ link function & $10^{\boldsymbol{\xi}}$ \\
                                       & \\
                \textit{Architecture} & Figure \ref{fig:unet} \\
                Recurrent steps ($T$) & 8 \\
                Number of parameters & $348\,546\,818$ \\
                                      & \\
                \textit{First Stage Optimisation} & \\
                Optimizer & Adamax \\
                Initial learning rate & $10^{-4}$\\
                Learning rate schedule & Exponential Decay \\
                Decay rate & 0.95 \\
                Decay steps & $100\,000$\\
                Number of steps & $610\,000$\\
                Batch size & 1 \\
                           & \\

                \textit{Second Stage Optimisation} & \\
                Optimizer & Adamax \\
                Initial learning rate & $6\times 10^{-5}$\\
                Learning rate schedule & Exponential Decay \\
                Decay rate & 0.9 \\
                Decay steps & $100\,000$\\
                Number of steps & $870\,000$\\
                Batch size & 1 \\
                
                \hline
        \end{tabular}
\end{table}


\section{Fine-Tuning}\label{ap:fine tuning}
We follow the work of \citet{Kirkpatrick2016} to define a prior distribution
over $\varphi$ that address the issue of catastrophic forgetting \citep{McCloskey1989,Ratcliff1990}:
\begin{equation}\label{eq:Prior} 
        \log p(\varphi) \propto -\frac{\lambda}{2}\sum_{j} \mathrm{diag}(\mathcal{I}(\varphi_{\mathcal{D}}^{\star}))_{j} 
        (\varphi_j - [\varphi^{\star}_{\mathcal{D}}]_{j})^{2},
\end{equation} 
where $\mathrm{diag}(\mathcal{I}(\varphi_{\mathcal{D}}^{\star}))$ is the diagonal of the 
Fisher information matrix 
encoding the amount of information that  
some set of gravitational lensing systems from 
the training set similar to the observed 
test task carries about the baseline RIM weights $\varphi_{\mathcal{D}}^{\star}$ --- the parameters that minimize the empirical risk over the training dataset $\mathcal{D}$.
The  
Lagrange multiplier $\lambda$ is 
tuning our estimated uncertainty about the neural network weights 
for the particular task at hand.  

We can understand the need for a conditional 
sampling distribution by looking at the posterior of the RIM parameters.
Suppose we are given a training set $\mathcal{D}$ and a test task $\mathcal{T}$ which are conditionally independent 
given $\varphi$ and have uniform priors, then the 
posterior of the RIM parameters $\mathcal{\varphi}$ can be rewritten using the Bayes rule as
\begin{equation}
        p(\varphi \mid \mathcal{D},\, \mathcal{T}) = 
\frac{p(\mathcal{T} \mid \varphi)  p(\varphi \mid \mathcal{D})}
        {p(\mathcal{T} \mid \mathcal{D})} \,.
\end{equation} 
The sampling distribution in this expression appears 
as the conditional $p(\mathcal{T} \mid \mathcal{D})$, which can also be viewed as 
the set of examples 
from the training set similar to the test task by rewriting it as 
$p(\mathcal{T} \mid \mathcal{D}) \propto p(\mathcal{D} \mid \mathcal{T})$.
The EWC term is then derived by a Laplace approximation of the prior $p(\varphi \mid \mathcal{D})$ around $\varphi_{\mathcal{D}}^{\star}$, which we also take to be 
proportional to the training loss, the likelihood of each time steps and an $\ell_2$ loss
\begin{equation}\label{eq:LossFisher}
        \log p\big(\varphi \mid (\mathbf{x}, \mathbf{y}) = \mathcal{D}\big) \propto -\mathcal{L}_{\varphi}(\mathbf{x}, \mathbf{y}) + \frac{1}{T}\sum_{t=1}^{T}\log p(\mathbf{y} \mid \mathbf{\hat{x}}^{(t)}) - \frac{\ell_2}{2}\lVert \varphi \rVert^2_2 \, .
\end{equation} 

Each reconstruction is performed by fine-tuning the baseline model 
on a test task composed of an observation vector, a PSF and a noise amplitude.
In practice, fine-tuning the test set of $3\,000$ examples can be accomplished in parallel so as to be done in 
at most a few days by spreading the computation on $\sim 10$ Nvidia A100 GPUs (or 10 hours on $\sim 100$ GPUs). 
Each reconstruction uses at most 2000 steps, which turns out to be approximately $20$ minutes (wall-time) per reconstruction. 
Early stopping is applied when the $\chi^2$ reaches noise level ($\chi^2 = \nu$).

\begin{minipage}{0.5\linewidth}
\begin{figure}[H]
        \centering
        \includegraphics[width=\linewidth]{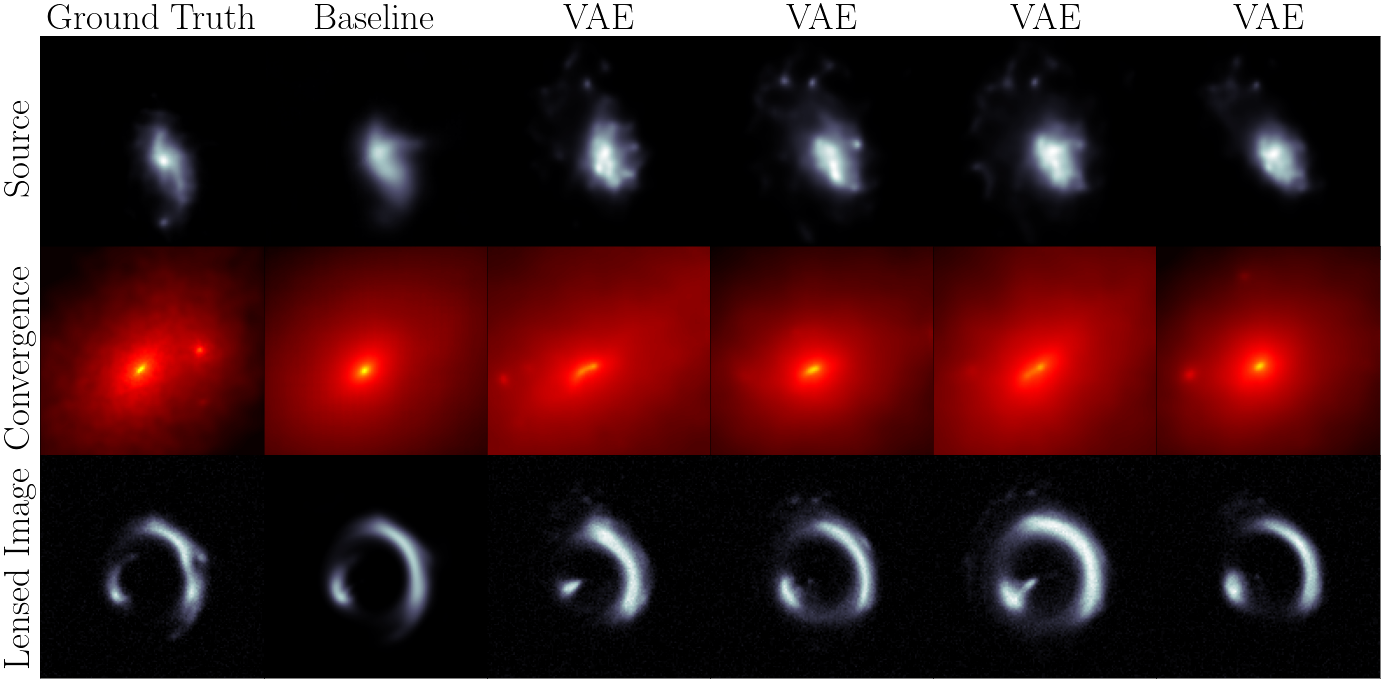}
        \caption{Examples similar to the test task shown in Figure \ref{fig:main result}. 
                They are sampled from the latent space of the source VAE and convergence VAE 
                near the RIM baseline latent  code $\mathbf{\hat{z}}^{(T)}$ and used to estimate $\mathrm{diag}(\mathcal{I}(\varphi^{\star}_{\mathcal{D}}))$.}
        \label{fig:vae fine-tuning}
\end{figure}
\end{minipage}\hfill
\begin{minipage}{0.5\linewidth}
\begin{table}[H]
        \centering
        \caption{Hyperparameters for fine-tuning the RIM.}
        \label{tab:fine-tuning hparams}
        \begin{tabular}{cc}
                Parameter & Value \\\hline\hline
                Optimizer & RMSProp \\
                Learning rate & $10^{-6}$\\
                Maximum number of steps & $2\,000$\\
                $\lambda$ & $2\times 10^{5}$\\
                $\ell_2$ & 0\\
                Number of samples from VAE & 200 \\
                Latent space distribution & $\mathcal{N}(\mathbf{\hat{z}}^{(T)}, \sigma=0.3)$\\
                \hline
        \end{tabular}
\end{table}
\end{minipage}
\vfill\null
\begin{figure}[H]
        \centering
        \begin{tikzpicture}]
                \tikzstyle{every node}=[font=\scriptsize]
                \node at (0, 0) {\includegraphics[width=\linewidth]{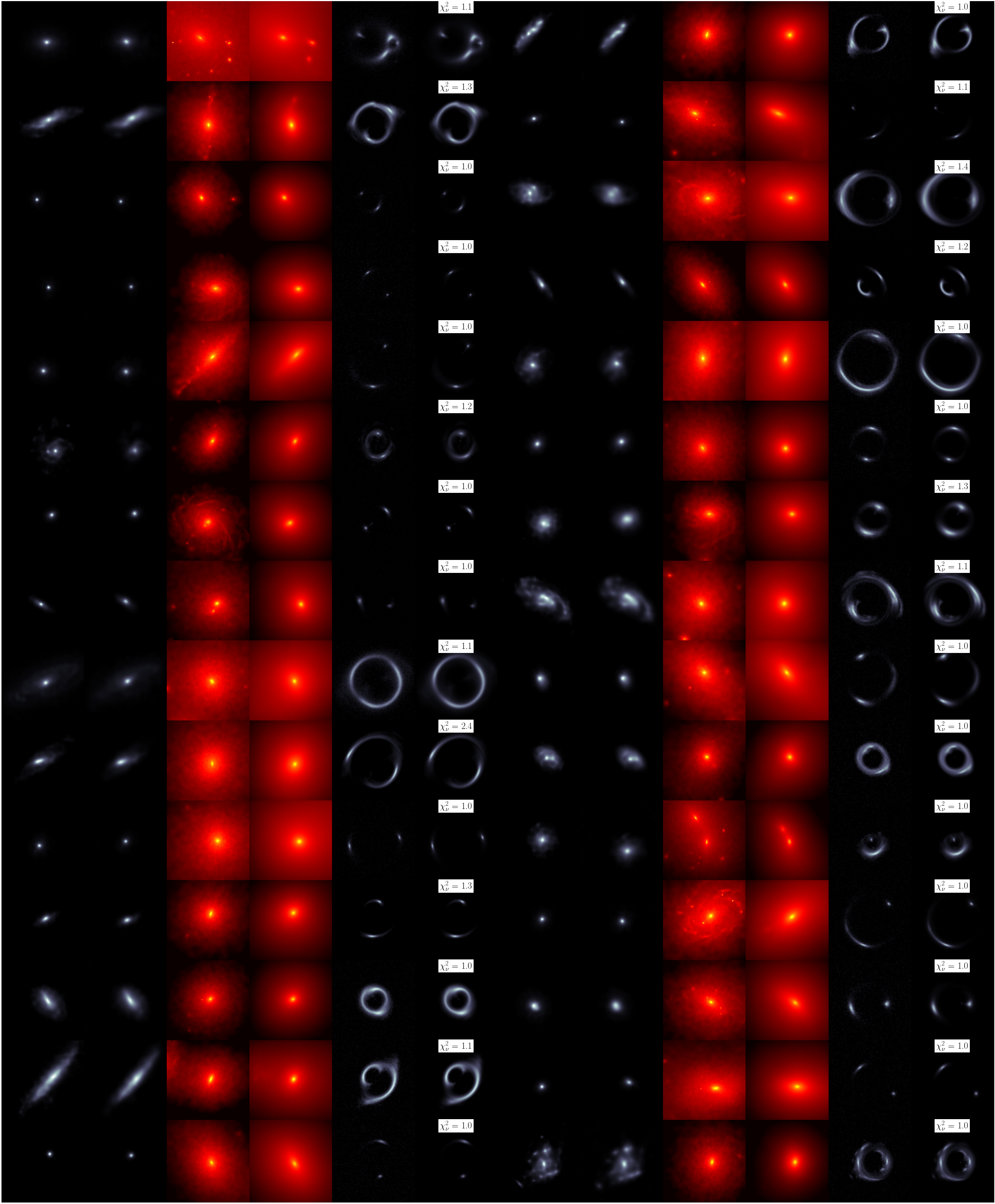}};
                \node at (-7.9, 10.5) {COSMOS};
                \node at (-6.5, 10.5) {RIM+FT};
                \node at (-5, 10.5) {IllustrisTNG};
                \node at (-3.6, 10.5) {RIM+FT};
                \node at (-2, 10.5) {Lensed Image};
                \node at (-0.6, 10.5) {RIM+FT};

                \node at (0.6, 10.5) {COSMOS};
                \node at (2, 10.5) {RIM+FT};
                \node at (3.6, 10.5) {IllustrisTNG};
                \node at (5, 10.5) {RIM+FT};
                \node at (6.5, 10.5) {Lensed Image};
                \node at (7.8, 10.5) {RIM+FT};

        \end{tikzpicture}
        \caption{
                30 reconstructions taken at random from the test set of 3000 examples simulated from COSMOS 
                and IllustrisTNG data at high SNR.
                The colorscale are the same as in Figure \ref{fig:main result}.}
        \label{fig:random sample}
\end{figure}


\end{document}